\documentclass[12pt]{article}
\usepackage{a4}
\usepackage{graphicx}
\usepackage{amsfonts}
\usepackage{amscd}
\usepackage{latexsym}
\usepackage{epsf}
\usepackage{amsmath}
\numberwithin{equation}{section}

\newcommand{\be}{\begin{equation}}
\newcommand{\ee}{\end{equation}}

\newcommand{\bea}{\begin{eqnarray}}
\newcommand{\eea}{\end{eqnarray}}

\newcommand{\f}{function }

\newcommand{\eqqs}{equations }

\newcommand{\nn}{\nonumber}

\newcommand{\fr}{\frac}

\newcommand{\pd}{\partial}

\newcommand{\ra}{\rightarrow}
\newcommand{\half}{\frac{1}{2}}

\begin{document}

\title{Two- and Three-dimensional Generalisation of Lower Order Local Wave Velocities}
\author{
{\sc I.~V.~Drozdov }\thanks{e-mail: drosdow@uni-koblenz.de} \\
{\small and}\\
A.~A.~Stahlhofen \thanks{e-mail: alfons@uni-koblenz.de}\\
\small  University of Koblenz, Institut f\"ur Naturwissenschaften\\
\small  Abteilung Physik\\
\small  Universit\"atsstr.1, D-56080 Koblenz, Germany\\
\it\small Language to be improved }
\maketitle

\begin{abstract}
 A general local approach for the definition of velocities
 and especially phase velocities for waves recently proposed
 for one-dimensional waves is generalized for 2 and 3 dimensional
 scalar wave.   A geometrically consistent generalization has been
  found for the local wave velocities of order zero and one.
\end{abstract}

\section{ Introduction }

 This paper is an extension of our recent work \cite{WV} being devoted to the definition and basic properties of local velocities spectrum of arbitrary
waves in 1+1 space-time dimensions. There it has been mentioned,
that the local concept proposed allows a straightforward and
unique generalization on an arbitrary dimension $N$, at least for
the physically most important cases of the ordinary zero order
phase velocity (0-PV or attribute velocity) and the first order
velocity (1-PV or peak velocity).

The formalism performed here is still developed for an object
defined by a distribution of some one-valued "attribute" in the
space-time as a single valued \f of coordinates $\{ x, t \}$ that
can be considered therefore as a scalar field. Hence, the
velocities determined below describe propagation of perturbations
of this field.

\section{ Velocity of zero-order (0-PV) }

Let the state of a medium be now described in 2-dimensional space
$(x,y)$ as a scalar
 field $\psi(x,y,t)$. As an illustration one can imagine a propagating perturbation on
a two-dimensional surface, e.g. a brane.

We follow the approach developed in Sec.2-4 of \cite{WV} and assume, the traced attribute is the certain fixed value
of the field: $\psi(x,y,t)=\psi_0=\mbox{const}$. This condition leads to
the requirement:
\be
d\psi\equiv\psi_x dx+ \psi_y dy+ \psi_t dt=0,
\ee
that means for velocity components $v^{(00)}$:

\be \psi_x \fr{dx}{dt}+\psi_x \fr{dy}{dt}= \psi_x v^{(00)}_x +
\psi_y v^{(00)}_y = -\psi_t \label{ansatz_00} \ee The lower
indices denote conventionally the partial derivative with respect
to the corresponding variable.

Since the forms of kind \be v^{(00)}_x=-C_x \fr{\psi_t}{\psi_x};\
v^{(00)}_y=-C_y \fr{\psi_t}{\psi_y} \label{def_00_C} \ee satisfy
Eq.(\ref{ansatz_00}) with arbitrary coefficients $C_x, C_y$
obeying the relation \be C_x+C_y=1\ee the definition
(\ref{def_00_C}) seems to be ambiguous. However, the coefficients
$C_x, C_y$ can be
 fixed through the meaningful geometrical requirement of transformation properties.
 Indeed, if we set $C_x=C_y=1/2$, we make sure that the doublet of reciprocal velocities:
 $ {\bf\tilde v}^{(00)}\equiv\{ 1/v^{(00)}_x; 1/ v^{(00)}_y \} $ transforms under coordinate
 change $(x,y) \ra (X,Y)$ as a covariant vector (co-vector) in the basis of the coordinate system.

 \be
 v^{(00)}_x=-\half\fr{\psi_t}{\psi_X X_x+\psi_Y Y_x},\ \ \ \ v^{(00)}_y=-\half\fr{\psi_t}{\psi_X X_y+\psi_Y Y_y}\\
 \ee

 that means:
 \be
 {\bf\tilde v}^{(00)}\equiv
 \left\{\begin{array}{c} 1/v^{(00)}_x\\  \\1/ v^{(00)}_y \end{array} \right\}=
\left[ \begin{array}{lr}  \fr{\pd X}{\pd x}\ \ \ \  & \fr{\pd Y}{\pd x}\\  \\ \fr{\pd X}{\pd y} \ \ \ \  & \fr{\pd
Y}{\pd y} \end{array}\right]
\left\{ \begin{array}{c} 1/v^{(00)}_X\\ \\1/ v^{(00)}_Y \end{array}  \right\}
 \ee

 Similarly, for the 0-PV components of three-dimensional scalar wave $\psi(x,y,z,t)$ (e.g. pressure in gas or liquid)
 we will obtain

\be
v^{(000)}_x=-\fr{1}{3} \fr{\psi_t}{\psi_x};\   v^{(000)}_y=-\fr{1}{3} \fr{\psi_t}{\psi_y};\  v^{(000)}_z=-\fr{1}{3}
\fr{\psi_t}{\psi_z}
\label{def_000}
 \ee
and the reciprocal velocity triplet $  {\bf\tilde v}^{(000)}\equiv \{ 1/v^{(000)}_x; 1/ v^{(000)}_y; 1/ v^{(000)}_z \}$
transforms as a 3-dimensional co-vector.

Generally, the components of N-dimensional 0-PV for scalar wave
$\psi(x_1,x_2,...,x_N, t)$ are defined as \be
v^{(N\times0)}_i=-\fr{1}{N} \fr{ {\pd \psi}/{\pd t}} {{\pd
\psi}/{\pd x_i}} \ee with the corresponding transformation
properties for $N$-multiplet of reciprocal components $
1/v^{(N\times0)}_i $ as an $N$-dimensional  co-vector.

\section{First-order velocity (1-PV) }

\subsection{Two-dimensional case}

Suppose, the traced point is labelled by an attribute of first order, i.e. a fixed value of derivative. To this end we fix
two values for spatial partial derivatives of $\psi$

\be
\{   \psi_x=C_1;\ \ \psi_y=C_2, \}
\ee

where $C_1,C_2- $ some constants. Especially it can be zero, if we trace a propagation of a positive/negative peak or a saddle point of the surface.

 \be
 d\left\{\begin{array}{c} \psi_x\\  \\  \psi_y \end{array} \right\}= 0 =
 \left\{\begin{array}{c} \psi_{xx}dx+\psi_{xy}dy+\psi_{xt}dt\\  \\  \psi_{xy}dx+\psi_{yy}dy+\psi_{yt}dt \end{array} \right\}=
 \ee

 This condition leads to the system of two \eqqs
  \bea
&&  \psi_{xx} v^{(11)}_x+ \psi_{xy} v^{(11)}_y = -\psi_{xt}\nn\\
&&  \psi_{xy} v^{(11)}_x+ \psi_{yy} v^{(11)}_y = -\psi_{yt},
  \eea
 and the concerning two components of 1-PV are the solutions:

  \be
  v^{(11)}_y=\fr{    \left|\begin{array}{cc} \psi_{xy} & \psi_{xt} \\ \psi_{yy} & \psi_{yt} \end{array} \right|      }
  {  \left|\begin{array}{cc}   \psi_{xx} & \psi_{xy} \\  \psi_{xy} & \psi_{yy}   \end{array} \right| }\equiv
  \fr{ \psi_{xy}\psi_{xt}-\psi_{xx}\psi_{yt}  }{ \psi_{xx}\psi_{yy}-\psi_{xy}^2 };\
   v^{(11)}_x=\fr{    \left|\begin{array}{cc} \psi_{xy} & \psi_{xt} \\ \psi_{yy} & \psi_{yt} \end{array} \right|      }
  {  \left|\begin{array}{cc}   \psi_{xx} & \psi_{xy} \\  \psi_{xy} & \psi_{yy}   \end{array} \right| }\equiv
  \fr{ \psi_{xy}\psi_{yt}-\psi_{yy}\psi_{xt}  }{ \psi_{xx}\psi_{yy}-\psi_{xy}^2 }.
  \ee

To recover a geometric nature of this doublet, we perform a transformation of coordinates as in the former case
$ (x,y) \ra (X,Y) $.

 Obviously, the transformation of ${\bf v}^{(11)}=\{  v^{(11)}_x,  v^{(11)}_y  \}$ is non-linear and involves second derivatives
 of coordinates as well as first derivatives of the \f $\psi$. In other words, the finite transformation law shows,
 that, for an arbitrary coordinate change, the doublet $ {\bf v}^{(11)} $ transforms not solely through its components itself, but
 also through the components of lower order velocity ${ \bf v^{(00)} }$.

 On account of this we consider the infinitesimal coordinate
transformation and respect only the linear part of it, it means,
we omit all the terms containing second derivatives of
coordinates. It arranges to establish the transformation law in
the form:

 \be
  \left\{\begin{array}{c}   v^{(11)}_x \\  \\   v^{(11)}_y  \end{array} \right\}=
  \left[ \fr{\pd (X,Y)}{\pd (x,y)}   \right]^{-1}  \left\{\begin{array}{c}   v^{(11)}_X \\  \\   v^{(11)}_Y  \end{array} \right\}
 \ee
 where the Jacobi-matrix:
 \be
 J= \left[ \fr{\pd (X,Y)}{\pd (x,y)}   \right]=
  \left[ \begin{array}{lr}  \fr{\pd X}{\pd x}\ \ \ \  & \fr{\pd X}{\pd y}\\  \\ \fr{\pd Y}{\pd x} \ \ \ \  & \fr{\pd
Y}{\pd y} \end{array}\right],
 \ee
 and we assured, that the doublet  ${\bf v}^{(11)}$ transforms as a true vector (contra-variant vector) in the basis of spatial coordinates $\{x,y\}$.
  It follows from the Sec.2, that the contraction product
    \be
   {\cal V}^{(11)}_{(00)} = {\bf\tilde v}^{(00)} \cdot  {\bf v}^{(11)}  \equiv   \fr{1}{v^{(00)}_x} v^{(11)}_x  + \fr{1}{v^{(00)}_y} v^{(11)}_y
    \ee
 is a dimensionless value and behaves as a true scalar under transformations of spatial coordinates.

   The both vectors respect the mirror transformations of the coordinates $(x,y) \ra  (-x,-y) $ as true vectors, as it is easy to see
   from their definitions.

\subsection{Three- and higher-dimension}

    A quite analogous procedure holds for the definition of 1-PV $v^{(111)}$ in a three-dimensional case. As an example
     a scalar field of potential $\psi(x,y,z)$ can be considered. The velocity of propagation of a gradient has to be defined.
      It follows from the condition:

     \be
 d\left\{\begin{array}{c} \psi_x\\  \\  \psi_y  \\  \\ \psi_z \end{array} \right\}=
 \left\{\begin{array}{c}
  \psi_{xx}dx+\psi_{xy}dy+ \psi_{xz}dz +  \psi_{xt}dt\\  \\  \psi_{xy}dx+\psi_{yy}dy+  \psi_{yz}dz + \psi_{yt}dt \\ \\
  \psi_{xz}dx+\psi_{yz}dy+ \psi_{zz}dz + \psi_{zt}dt \end{array} \right\}=0
 \ee

  With the determinants of this system:
  \be
  \small
     D = \left|\begin{array}{ccc}
     \psi_{xx} & \psi_{xy}  &  \psi_{xz} \\
       \psi_{xy} & \psi_{yy}  &  \psi_{yz} \\
        \psi_{xz} & \psi_{yz}  &  \psi_{zz}
      \end{array} \right|,
     D_x =\left|\begin{array}{ccc}
     -\psi_{xt} & \psi_{xy}  &  \psi_{xz} \\
       -\psi_{yt} & \psi_{yy}  &  \psi_{yz} \\
        -\psi_{zt} & \psi_{yz}  &  \psi_{zz}
      \end{array} \right|,
     D_y =\left|\begin{array}{ccc}
     \psi_{xx} & -\psi_{xt}  &  \psi_{xz} \\
       \psi_{xy} & -\psi_{yt}  &  \psi_{yz} \\
        \psi_{xz} & -\psi_{zt}  &  \psi_{zz}
      \end{array} \right|,
     D_z =\left|\begin{array}{ccc}
     \psi_{xx} & \psi_{xy}  &  -\psi_{xt} \\
       \psi_{xy} & \psi_{yy}  &  -\psi_{yt} \\
        \psi_{xz} & \psi_{yz}  &  -\psi_{zt}
      \end{array} \right|,
  \ee
   \normalsize
     the components of the 3-dimensional 1-PV vector are defined as:

  \be
   v^{(111)}_x =\fr{D_x}{D},\ \       v^{(111)}_y =\fr{D_y}{D},\ \       v^{(111)}_z =\fr{D_z}{D}
   \ee
  and it follows for the dimensionless scalar velocity
  \be
   {\cal V}^{(111)}_{(000)} = {\bf\tilde v}^{(000)} \cdot  {\bf v}^{(111)}  \equiv   \fr{1}{v^{(000)}_x} v^{(111)}_x  +
   \fr{1}{v^{(000)}_y} v^{(111)}_y +  \fr{1}{v^{(000)}_z} v^{(111)}_z
    \ee
    respectively.

   The similar analysis can be performed  and remains to be valid for an arbitrary space dimension $N$.

\end{document}